\documentclass[twoside,twocolumn,9pt]{article}
\usepackage{extsizes}
\usepackage[super,sort&compress,comma]{natbib}
\usepackage[version=3]{mhchem}
\usepackage[left=1.5cm, right=1.5cm, top=1.785cm, bottom=2.0cm]{geometry}
\usepackage{balance}
\usepackage{times,mathptmx}
\usepackage{sectsty}
\usepackage{graphicx}
\usepackage{lastpage}
\usepackage[format=plain,justification=justified,singlelinecheck=false,font={stretch=1.125,small,sf},labelfont=bf,labelsep=space]{caption}
\usepackage{float}
\usepackage{fancyhdr}
\usepackage{fnpos}
\usepackage[english]{babel}
\usepackage{array}
\usepackage{droidsans}
\usepackage{charter}
\usepackage[T1]{fontenc}
\usepackage[usenames,dvipsnames]{xcolor}
\usepackage{setspace}
\usepackage[compact]{titlesec}
\usepackage{hyperref}

\usepackage{epstopdf}

\definecolor{cream}{RGB}{222,217,201}

\begin{document}

\pagestyle{fancy}
\thispagestyle{plain}
\fancypagestyle{plain}{

\fancyhead[C]{\includegraphics[width=18.5cm]{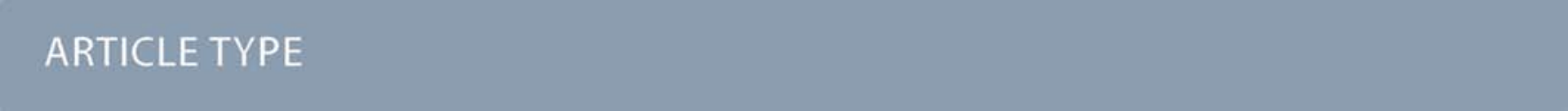}}
\fancyhead[L]{\hspace{0cm}\vspace{1.5cm}\includegraphics[height=30pt]{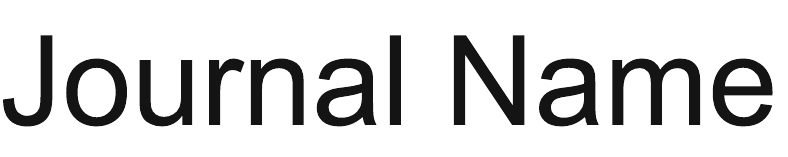}}
\fancyhead[R]{\hspace{0cm}\vspace{1.7cm}\includegraphics[height=55pt]{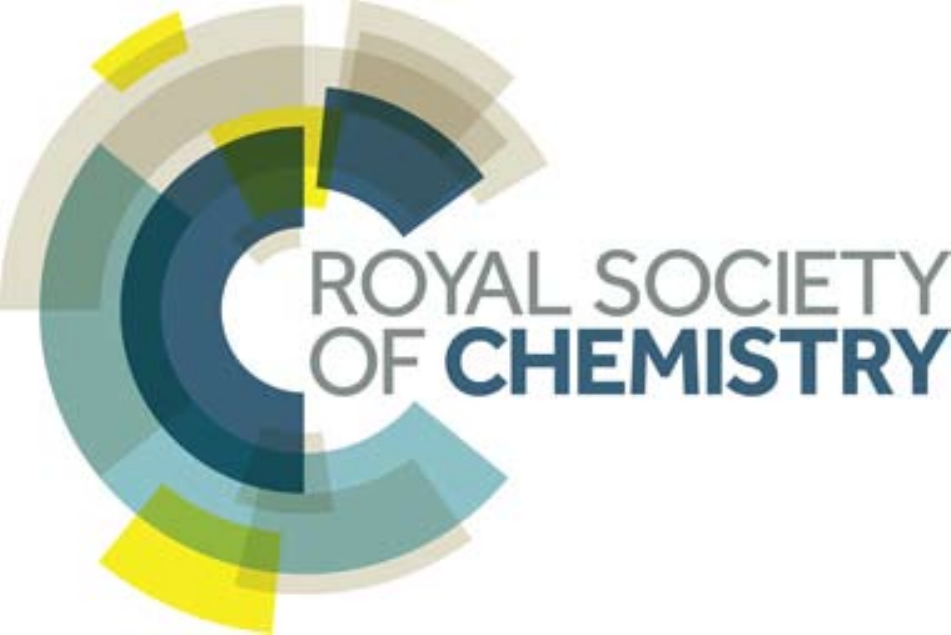}}
\renewcommand{\headrulewidth}{0pt}
}

\makeFNbottom
\makeatletter
\renewcommand\LARGE{\@setfontsize\LARGE{15pt}{17}}
\renewcommand\Large{\@setfontsize\Large{12pt}{14}}
\renewcommand\large{\@setfontsize\large{10pt}{12}}
\renewcommand\footnotesize{\@setfontsize\footnotesize{7pt}{10}}
\renewcommand\scriptsize{\@setfontsize\scriptsize{7pt}{7}}
\makeatother

\renewcommand{\thefootnote}{\fnsymbol{footnote}}
\renewcommand\footnoterule{\vspace*{1pt}%
\color{cream}\hrule width 3.5in height 0.4pt \color{black} \vspace*{5pt}}
\setcounter{secnumdepth}{5}

\makeatletter
\renewcommand\@biblabel[1]{#1}
\renewcommand\@makefntext[1]%
{\noindent\makebox[0pt][r]{\@thefnmark\,}#1}
\makeatother
\renewcommand{\figurename}{\small{Fig.}~}
\sectionfont{\sffamily\Large}
\subsectionfont{\normalsize}
\subsubsectionfont{\bf}
\setstretch{1.125} 
\setlength{\skip\footins}{0.8cm}
\setlength{\footnotesep}{0.25cm}
\setlength{\jot}{10pt}
\titlespacing*{\section}{0pt}{4pt}{4pt}
\titlespacing*{\subsection}{0pt}{15pt}{1pt}

\fancyfoot{}
\fancyfoot[LO,RE]{\vspace{-7.1pt}\includegraphics[height=9pt]{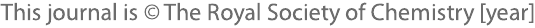}}
\fancyfoot[CO]{\vspace{-7.1pt}\hspace{13.2cm}\includegraphics{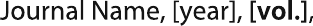}}
\fancyfoot[CE]{\vspace{-7.2pt}\hspace{-14.2cm}\includegraphics{head_foot/RF}}
\fancyfoot[RO]{\footnotesize{\sffamily{1--\pageref{LastPage} ~\textbar  \hspace{2pt}\thepage}}}
\fancyfoot[LE]{\footnotesize{\sffamily{\thepage~\textbar\hspace{3.45cm} 1--\pageref{LastPage}}}}
\fancyhead{}
\renewcommand{\headrulewidth}{0pt}
\renewcommand{\footrulewidth}{0pt}
\setlength{\arrayrulewidth}{1pt}
\setlength{\columnsep}{6.5mm}
\setlength\bibsep{1pt}

\makeatletter
\newlength{\figrulesep}
\setlength{\figrulesep}{0.5\textfloatsep}

\newcommand{\topfigrule}{\vspace*{-1pt}%
\noindent{\color{cream}\rule[-\figrulesep]{\columnwidth}{1.5pt}} }

\newcommand{\botfigrule}{\vspace*{-2pt}%
\noindent{\color{cream}\rule[\figrulesep]{\columnwidth}{1.5pt}} }

\newcommand{\dblfigrule}{\vspace*{-1pt}%
\noindent{\color{cream}\rule[-\figrulesep]{\textwidth}{1.5pt}} }

\makeatother

\twocolumn[
  \begin{@twocolumnfalse}
\vspace{3cm}
\sffamily
\begin{tabular}{m{4.5cm} p{13.5cm} }

\includegraphics{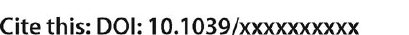} & \noindent\LARGE{Dynamical process of optically trapped singlet ground state $^{85}$Rb$^{133}$Cs molecules produced via short-range photoassociation} \\
 & \vspace{0.3cm} \\
 & \noindent\large{Zhonghao Li,\textit{$^{a,b}$} Ting Gong,\textit{$^{a,b}$} Zhonghua Ji,\textit{$^{a,b}$} Yanting Zhao,$^{\ast}$\textit{$^{a,b}$} Liantuan Xiao,\textit{$^{a,b}$} and Suotang Jia\textit{$^{a,b}$}} \\ & \vspace{-0.3cm} \\
\includegraphics{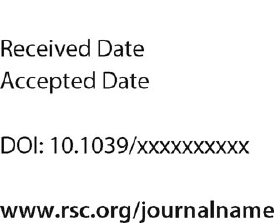} &\noindent\sffamily{We investigate the dynamical process of optically trapped X$^{1}$$\Sigma$$^{+}$ (v'' = 0) state $^{85}$Rb$^{133}$Cs molecules distributing in J'' = 1 and J'' = 3 rotational states. The considered molecules, formed from short-range photoassociation of mixed cold atoms, are subsequently confined in a crossed optical dipole trap. Based on a phenomenological rate equation, we provide a detailed study of the dynamics of $^{85}$Rb$^{133}$Cs molecules during the loading and holding processes. The inelastic collisions of $^{85}$Rb$^{133}$Cs molecules in the X$^{1}$$\Sigma$$^{+}$ (v'' = 0, J'' = 1 and J'' = 3) states with ultracold $^{85}$Rb (or $^{133}$Cs) atoms are measured to be 1.0 (2)$\times$10$^{-10}$ cm$^{3}$s$^{-1}$ (1.2 (3)$ \times$ 10$^{-10}$ cm$^{3}$s$^{-1}$). Our work provides a simple and generic procedure for studying the dynamical process of trapped cold molecules in the singlet ground states.}\\ &\vspace{0.3cm}
 \\
\end{tabular}

 \end{@twocolumnfalse} \vspace{0.6cm}

  ]

\renewcommand*\rmdefault{bch}\normalfont\upshape
\rmfamily
\section*{}
\vspace{-1cm}


\footnotetext{\textit{$^{a}$Shanxi University, State Key Laboratory of Quantum Optics and Quantum Optics Devices, Institute of Laser Spectroscopy, Wucheng Rd. 92,  030006 Taiyuan, China. \\E-mail: zhaoyt@sxu.edu.cn}}
\footnotetext{\textit{$^{b}$Shanxi University, Collaborative Innovation Center of Extreme Optics, Wucheng Rd. 92, 030006 Taiyuan, China.}}

\rmfamily 

\section*{Introduction}

The~preparation~and~manipulation~of~ultracold molecules have attracted significant interests in the past few years.~Due to strong dipole-dipole interactions, ultracold heteronuclear molecules in the absolute ground state are particularly attractive: With permanent electric dipole moments (EDMs) that ranges from a half to several Debye~{\color{blue}\cite{AymarDulieu2005}}, such molecules could interact strongly with external fields, as well as with each other via long-range dipole-dipole force. These novel molecular properties promise interesting applications in the precision measurement ~{\color{blue}\cite{ZelevinskyKotochigovaYe2008,Schiller2007,DeMilleCahnMurphreeEtAl2008}}, quantum control of cold chemical reaction ~{\color{blue}\cite{JoopJ2006,Ospelkaus2010}}, and quantum computation~{\color{blue}\cite{DeMille2002}}.

At present, there exist various proposals on the preparation of ground-state polar molecules, in particular those in the lowest rovibrational ground state. In this context, one of the most promising approach involves combination of ~the magnetoassociation (MA) with stimulated Raman adiabatic passage (STIRAP), in sense of achievable temperature and phase space density. This has led to the achievement of $^{40}$K$^{87}$Rb~{\color{blue}\cite{OspelkausNiQuemenerEtAl2010}}, $^{87}$Rb$^{133}$Cs~{\color{blue}\cite{MolonyGregoryJiEtAl2014,Takekoshi2014}}, $^{23}$Na$^{40}$K~{\color{blue}\cite{ParkWillZwierlein2015}} and $^{23}$Na$^{87}$Rb~{\color{blue}\cite{GuoZhuLuEtAl2016}}.~However, this scheme is only suitable for atomic species where the MA is available and requires the initial cold atomic state to be nearly degenerate. Alternatively, photoassociation (PA) as a simple and universal method has been extensively applied. For example, combination of PA with ``pump-dump'' has been applied to $^{85}$Rb$^{133}$Cs~{\color{blue}\cite{SageSainisBergemanEtAl2005}}. Nevertheless, the transfer efficiency is low due to the large branching ratio to other electronic states and complicated optical pathways.~The PA has also been combined with STIRAP in~$^{41}$K$^{87}$Rb~{\color{blue}\cite{AikawaAkamatsuHayashiEtAl2010}}. While this give rise to a transfer process that is highly efficient and state-selective, the ground-state molecules can only be formed once per experimental cycle, thus limiting the accumulation of molecules. A third approach uses direct short-range PA, where the direct spontaneous emission after PA allows creation of molecules in the singlet ground state. Such scheme has been successfully implemented in $^{39}$K$^{85}$Rb~{\color{blue}\cite{BanerjeeRahmlowCarolloEtAl2012}}, $^{23}$Na$^{133}$Cs~{\color{blue}\cite{ZabawaWakimHaruzaEtAl2011}}, $^{7}$Li$^{133}$Cs~{\color{blue}\cite{DeiglmayrGrocholaReppEtAl2008}}, $^{85}$Rb$^{133}$Cs~{\color{blue}\cite{BruzewiczGustavssonShimasakiEtAl2014}} and $^{7}$Li$^{85}$Rb~{\color{blue}\cite{StevensonBlasingChenEtAl2016}}{\textit{et al.}.~Remarkably, the continuous production and simultaneous trapping of molecules via this method has the promising potential to produce a large number of molecules in special molecular states by optical pumping, such as vibrational cooling~{\color{blue}\cite{Matthieu2008}} and rotational cooling~{\color{blue}\cite{ManaiHorchaniLignierEtAl2012}}. These samples provide the basis to form pair-supersolid phase~{\color{blue}\cite{TrefzgerMenottiLewenstein2009}} and the molecular Bose-Einstein condensates~{\color{blue}\cite{GreinerRegalJin2003,BartensteinAltmeyerRiedlEtAl2004}}.

In addition, there have been significant efforts in confining molecules using various types of traps, such as electrostatic trap~{\color{blue}\cite{Hendrick2000}}, magnetic trap~{\color{blue}\cite{FriedrichWeinsteinEtAl1999}}, and optical dipole trap (ODT)~{\color{blue}\cite{ScotCKuo1993}} or their combinations.~The former two traps only apply for particular molecules, such as the electrostatic trap for the polar molecules and the magnetic trap for the paramagnetic molecules in the low-field-seeking states.~By contrast, the ODT is more widely used.~Formation of molecules via PA and confinement in an ODT have been demonstrated. For example, quasielectrostatic trap (QUEST) has been used to confine $^{133}$Cs$_{2}$~{\color{blue}\cite{ZahzamVogtMudrichEtAl2006,StaanumKraftLangeEtAl2006a}}, $^{87}$Rb$_{2}$~{\color{blue}\cite{Gabbanini2007}}, $^{85}$Rb$^{133}$Cs~{\color{blue}\cite{HudsonGilfoyKotochigovaEtAl2008}}, and $^{7}$Li$^{133}$Cs~{\color{blue}\cite{DeiglmayrReppDulieuEtAl2011,DeiglmayrReppWesterEtAl2011}}, while the far off-resonance optical dipole trap (FORT) has been applied to $^{87}$Rb$_{2}$ and $^{85}$Rb$_{2}$~{\color{blue}\cite{MenegattiMarangoniMarcassa2011,ChenKaoChenEtAl2013}}.~These trapped samples have been used for the calculation of the atom-molecule inelastic collision rate coefficients~{\color{blue}\cite{ZahzamVogtMudrichEtAl2006,StaanumKraftLangeEtAl2006a,DeiglmayrReppDulieuEtAl2011,DeiglmayrReppWesterEtAl2011}}, verification of the data acquisition technique~{\color{blue}\cite{MenegattiMarangoniMarcassa2011}}, and determination of the molecule number ~{\color{blue}\cite{ChenKaoChenEtAl2013}}. Motivated by the significant interests in the lowest vibrational X$^{1}\Sigma^{+}$ (v'' = 0) state, it is desirable to seek an effective production and storage of these ground state molecules.~Unlike above trapped molecular samples which are either in the a$^{3}\Sigma^{+}$ states or high-lying levels of X$^{1}\Sigma^{+}$ states, in this work we aim at creation and confinement of molecules in the X$^{1}$$\Sigma$$^{+}$ (v'' = 0) state.

\begin{figure*}[!htb]
\centering \includegraphics[angle=360,width=0.82\textwidth]{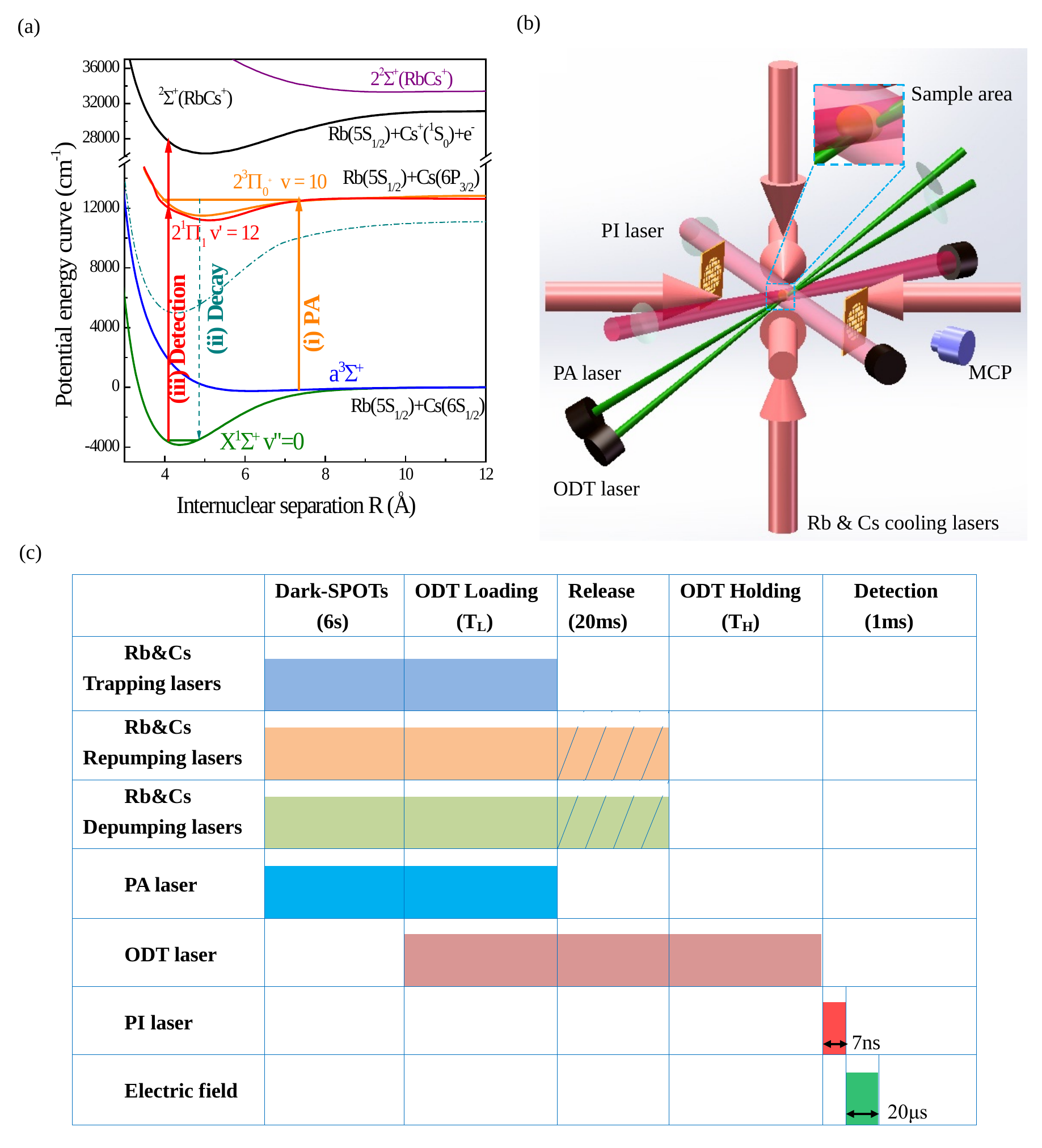}
\caption{(a) Formation and detection mechanism for ultracold $^{85}$Rb$^{133}$Cs molecules in the X$^{1}$$\Sigma$$^{+}$ (v'' = 0) state. The potential energy curves are based on the data in ref.~{\color{blue}\cite{FahsAlloucheKorekEtAl2002}}. The related pathways are the same with ref.\cite{BruzewiczGustavssonShimasakiEtAl2014}. (b) Brief diagram of optical elements near vacuum chamber. (c) Experimental time sequence. The shadow with different color means ``on'', the blank means ``off'', and the rectangle with lines means ``selectively on or off''.  When the lasers are kept on in order to push the co-trapped cold atoms out of the ODT while the lasers are turned off in order to keep the trapped cold atoms in the ODT. }
\end{figure*}

For $^{85}$Rb$^{133}$Cs molecules, the detailed information of the 2$^{1}$$\Pi$$_{1}$, 2$^{3}$$\Pi$$_{0^{-}}$, 2$^{3}$$\Pi$$_{0^{+}}$, 2$^{3}$$\Pi$$_{1}$, 3$^{3}$$\Sigma$$^{+}_{1}$ excited states via short-range PA have been reported~{\color{blue}\cite{Gabbanini2011,YuanZhaoJiEtAl2015,ZhaoYuanJiEtAl2016,ShimasakiKimDeMille2016}}, which give abundant choice with an effectively intermediate state for the preparation of  X$^{1}$$\Sigma$$^{+}$ (v'' = 0) state molecules.~Specially, C. D. Bruzewicz and co-workers have studied the continuous formation of ultracold $^{85}$Rb$^{133}$Cs molecules in X$^{1}$$\Sigma$$^{+}$ (v'' = 0 - 10) via short-range PA to the 2$^{3}$$\Pi$$_{0^{+}}$ (v = 10) intermediate state~{\color{blue}\cite{BruzewiczGustavssonShimasakiEtAl2014}}. These results provide powerful basis for the realization of optical trapping of  X$^{1}$$\Sigma$$^{+}$ (v'' = 0) state molecules via short-range PA. Although these molecules in X$^{1}$$\Sigma$$^{+}$ (v'' = 0) state have been produced via short-range PA, experimental achievement of the optical trapping has not yet been reported so far.

In this paper, ultracold $^{85}$Rb$^{133}$Cs molecules in the lowest vibrational ground state are confined in a crossed ODT. The ground state molecules are prepared via short-range PA and measured~selectively through the sensitive photoionization (PI) detection technology. A phenomenological rate equation is introduced to give the quantitative analysis of the loading and holding procedures of X$^{1}$$\Sigma$$^{+}$ (v'' = 0) state (to be precise, the J'' = 1 and J'' = 3 states of X$^{1}$$\Sigma$$^{+}$ (v'' = 0)) molecules in ODT. It is found that the inelastic molecular collision and the vibrational redistribution are negligible.~In addition, the inelastic collision of $^{85}$Rb$^{133}$Cs molecules in the X$^{1}$$\Sigma$$^{+}$ (v'' = 0, J'' = 1 and J'' = 3) states with ultracold $^{85}$Rb (or $^{133}$Cs) atoms is measured.

\section*{Experimental setup}

Figure 1(a) shows the diagram illustrating the formation and detection mechanism of ultracold $^{85}$Rb$^{133}$Cs molecules in the lowest vibrational ground state. (i) PA. Pairs of colliding $^{85}$Rb and $^{133}$Cs atoms are excited to a deeply bound state at short-range. (ii) Decay. After two-photon-cascade spontaneous emission process~{\color{blue}\cite{Shimasaki2015}}, the stable state molecules are distributing in different vibrational states~{\color{blue}\cite{BruzewiczGustavssonShimasakiEtAl2014}}. (iii) Detection. The molecules are detected by PI technology.

As shown in Fig. 1(b), the ultracold $^{85}$Rb and $^{133}$Cs atoms are initially cooled and trapped in overlapped dual species magneto-optical traps (MOTs). The MOTs are operated in dark spontaneous force optical traps (dark-SPOTs) configuration, which provide the atomic sample with high atomic density and low collision rate~{\color{blue}\cite{Ketterle1993}}.~The vacuum background pressure in the chamber is kept at about 3 $\times$ 10$^{-7}$ Pa.~A pair of anti-Helmholtz coils generates a magnetic gradient of about 15 G/cm.~Four Littrow external-cavity diode lasers (Toptica, DL100) locked by saturation absorption spectroscopy technique provide the trapping and repumping beams for $^{85}$Rb and $^{133}$Cs. All of these beams are 15 mm in diameter. The total power of trapping laser is around 36 mW, and the total power of repumping laser is around 8 mW, respectively. The dark-SPOTs are achieved by filling the depumping beams in the dark region of the repumping beams. The dark region is created by a black dot in the center of the mixed repumping beams. Both depumping beams have a power of about 80 $\mu$W with 3 mm diameter. The overlap of atomic clouds is verified by two charge-coupled device (CCD) cameras placed along the horizontal and vertical directions, respectively. In this way,  about 1 $\times$ 10$^{7}$ of $^{85}$Rb atoms are formed in the 5S$_{1/2}$ (F = 2) state with a density of 8 $\times$ 10$^{10}$ cm$^{-3}$, and about 2 $\times$ 10$^{7}$ $^{133}$Cs atoms are formed in the 6S$_{1/2}$ (F = 3) state with a density of 1.5 $\times$ 10$^{11}$ cm$^{-3}$.

The PA process is achieved by employing a tunable Ti: sapphire laser system (M Squared, Sols Ti:sapphire) with a typical linewidth of 100 kHz and an output power of up to 1.5 W. We focus the PA beam on the center of the overlapped dark-SPOTs with the Gaussian radius of 150 $\mu$m. The PA laser is locked by transfer cavity technique based on an ultrastable He-Ne laser as a reference laser. The produced molecules will subsequently decay to the ground state. The frequency is monitored using a commercial wavelength meter (High Finesse-Angstrom, WS-7R) with an absolute accuracy of 0.002 cm$^{-1}$ (about 60 MHz), which is calibrated with Rb atomic transition line.

The PI laser pathways as shown in Fig. 1(a) are chosen to observe the produced ground state molecules. The PI laser is provided by only a tunable dye laser (CBR-G-18EG, Spectra physics, wavelength is about 651.8 nm, pulse duration of 7 ns, diameter is about 3 mm, pulse energy is about 1 mJ, repetition rate of 10 Hz). The dye laser is pumped by the second harmonic of an Nd-YAG laser (Spectra physics, INDIE-40-10-HG, wavelength of 532 nm) and operated on DCM dissolved in dimethyl sulfoxide (DMSO). The ions formed in this process are accelerated by a pulsed electric field with a duration time of 20 $\mu$s. After the accelerated procedure, these moving ions will fly freely with length of 68 mm. Finally, the ions are detected by a pair of microchannel plates (MCPs). The electric signals are detected, amplified, then monitored on a digital oscilloscope and recorded using an NI PCI-1714 card following a boxcar (Boxcar, SRS-250) with 10 averages.

An ODT is built to trap the formed molecules. The simplest configuration of ODT is a single focused beam, which creates a highly anisotropic trap with relatively weak confinement along the propagation axis and tight confinement in the perpendicular direction. By crossing two focused beams, one can create a nearly isotropic and tight confinement in all dimensions, allowing formation of denser samples. In our experiment, the ODT is realized with a linearly polarized broadband fiber laser (IPG Photonics, YLR-300-AC) with central wavelength of 1070 nm and linewidth of 2 nm. The IPG laser is divided into two beams with orthogonal linear polarization. As shown in the Fig. 1(b), two ODT beams are crossed on the center of the dark-SPOTs with an angle of 45$^{o}$. The waists of these two beams are both 76 (2) $\mu$m. The frequencies are shifted oppositely by 110 MHz by using two acousto-optic modulators (AOMs) to prevent DC interference effects and allow rapid control of the ODT ($\tau<1\mu$s). The AC Stark shift produces a conservative potential with a minimum at the focus, where the ultracold atomic and molecular sample can be trapped. We find it is advantageous to move the point of the crossed ODT below the center of dark-SPOTs by 0.5 mm along the direction of gravity. The total power of ODT beams is normally around 5 W. About 8 $\times$ 10$^{5}$ of $^{85}$Rb atoms at the 5S$_{1/2}$ (F = 2) state with a density of 2 $\times$ 10$^{11}$ cm$^{-3}$ and about 1.5 $\times$ 10$^{6}$ $^{133}$Cs atoms at the 6S$_{1/2}$ (F = 3) state with a density of 4 $\times$ 10$^{11}$ cm$^{-3}$ are transferred from the dark-SPOTs.

The time sequence of the experiment is shown in Fig. 1(c), which is divided to five stages. 1) The loading procedure of dark-SPOTs. The PA procedure occurs simultaneously and the duration time is about 6 s. All the dark-SPOTs lasers are turned on in the presence of magnetic field to trap cold atoms. The cooling and repumping lasers are chosen at the optimized value for the largest loading rate. 2) The loading procedure of ODT. 3) The expansion procedure of untrapped atoms and molecules with a duration time of 20 ms. In this procedure, the depumping and repumping lasers can be selected to keep on as the pushing lasers to remove the mixed atoms in ODT or are turned off for the investigation of atom-molecule collisions. 4) The holding procedure of trapped samples in ODT. In this period, only the dipole laser is kept on. 5) Detection. The trapped molecules are detected by PI through the resonance-enhanced two-photon ionization (RETPI).

\section*{Experimental results and analysis}

\subsection*{Optical trapping of X$^{1}$$\Sigma$$^{+}$ (v'' = 0) state molecules distributing in J'' = 1 and J'' = 3 rotational states}

Figure 2(a) shows the rotational structure of the excited 2$^{3}$$\Pi$$_{0^{+}}$ (v = 10) state, which agree with the measured results in ref. \cite{BruzewiczGustavssonShimasakiEtAl2014}. ~Because this state has large free-to-bound and bound-to-bound Frank-Condon factors (FCFs),~it can be used for the preparation of molecules in X$^{1}$$\Sigma$$^{+}$ (v'' = 0) state with the ratio of about 30\% efficiently.~In the following experiments, the frequency of the PA laser is fixed at J = 1. The rotational distribution of molecules in X$^{1}$$\Sigma$$^{+}$ (v'' = 0) state are J'' = 1 and J'' = 3 according to ref.~\cite{Shimasaki2015} and also have been confirmed in our measurement by depletion spectroscopy. To be concise, X$^{1}$$\Sigma$$^{+}$ (v'' = 0) represents X$^{1}$$\Sigma$$^{+}$ (v'' = 0, J'' = 1 and J'' = 3) states in the following contexts.

The X$^{1}$$\Sigma$$^{+}$ (v'' = 0) state molecules could be state-selectively detected via RETPI. The PI spectra have been measured by the scanning PI laser frequency (about 630nm-680nm) while keeping the PA laser fixed at 2$^{3}$$\Pi$$_{0^{+}}$ (v = 10, J = 1) level. Here we show part of PI spectra of X$^{1}$$\Sigma$$^{+}$ (v'' = 0) state in Fig. 2(b). Since the frequency bandwidth of PI laser (about 0.2 cm$^{-1}$) is larger than the energy spacing of rotational level, it is impossible to distinguish rotational levels. The spectra allow for straigtforward assignment of the observed transitions between vibrational states based on ref. \cite{BruzewiczGustavssonShimasakiEtAl2014} and ref.~\cite{LeeYoonLeeEtAl2008}. We assign the  $\sim$ 15278.31 cm$^{-1}$ peak to the transition 2$^{1}$$\Pi$$_{1}$ (v' = 10) $\leftarrow$ X$^{1}$$\Sigma$$^{+}$ (v'' = 0)  and $\sim$ 15341.41 cm$^{-1}$ peak to 2$^{1}$$\Pi$$_{1}$ (v' = 12) $\leftarrow$ X$^{1}$$\Sigma$$^{+}$ (v'' = 0) transition. The red line is the Lorentzian fitting.  In following work, the PI laser frequency is fixed at about 15341.41 cm$^{-1}$.

Based on the sensitive PI technology, the molecules in the X$^{1}$$\Sigma$$^{+}$ (v'' = 0) state can be compared directly with the time of flight (TOF) mass spectrum in two conditions: the ``ODT off'' and ``ODT on''. The ``ODT off'' means the molecules are released free after regular PA procedure without ODT and then detected, while the ``ODT on'' means the molecules are transferred to ODT and then detected. In order to obtain obvious contrast, we record the TOF mass spectrum when the PA laser and all cooling lasers are turned off 20 ms before the detection procedure in these two conditions. With the data in Fig. 2(c), the molecular ion signal shows 5 times enhanced in``ODT on'' compared with the ``ODT off''.~These results show that the $^{85}$Rb$^{133}$Cs molecules in the X$^{1}$$\Sigma$$^{+}$ (v'' = 0) state have been trapped in the ODT. In addition, the \textit{$\Delta$m/z} is 1.18 by fitting the data with Lorentzian function. Considering the \textit{m/z} of  $^{85}$Rb$^{133}$Cs$^{+}$ is 218, the resolution of the TOF mass spectrometer can be found as 185.

\begin{figure}[H]
  \centerline{\includegraphics[width=0.8\columnwidth]{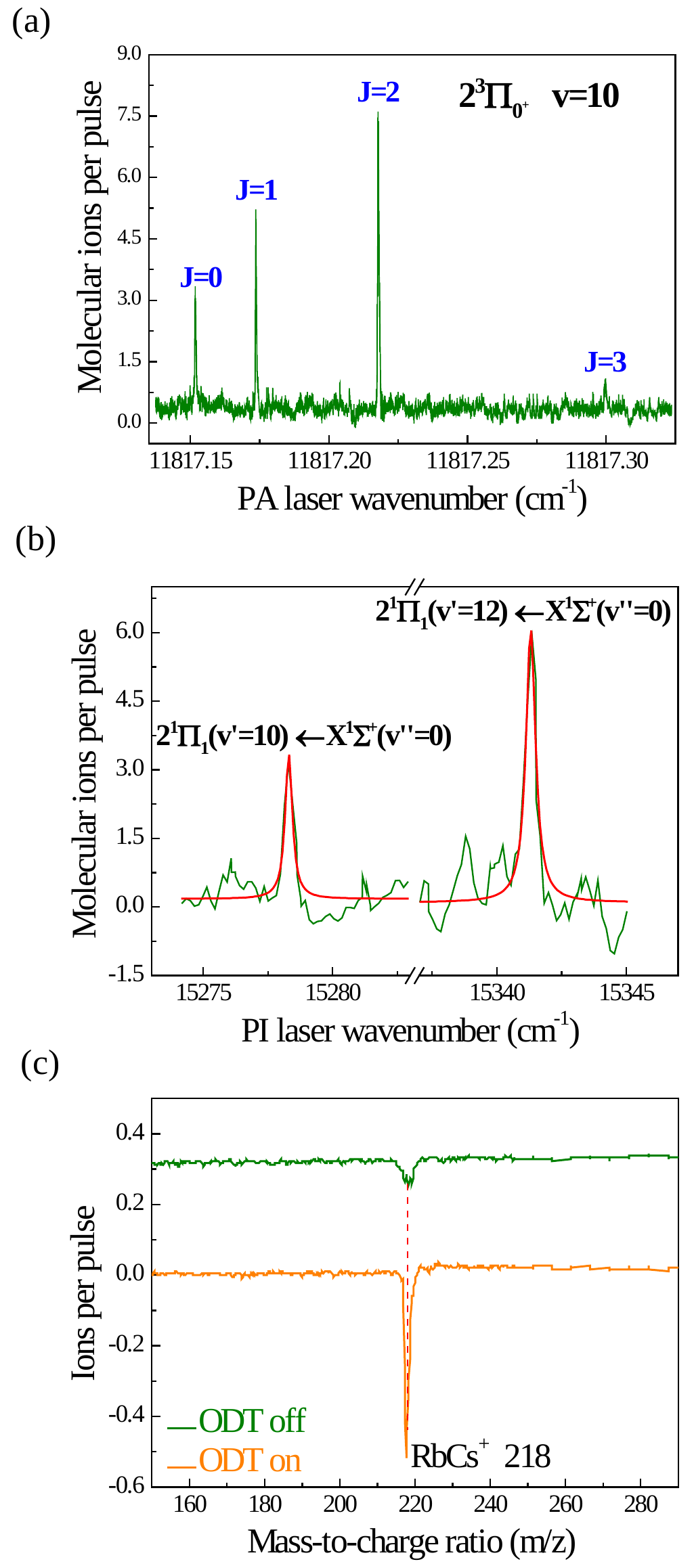}}
  \caption{(a) PA spectrum of 2$^{3}$$\Pi$$_{0^{+}}$ (v = 10) state, which is the intermediate state for the preparation of $^{85}$Rb$^{133}$Cs molecules in the X$^{1}$$\Sigma$$^{+}$ (v'' = 0) state. In the following experiments, the frequency of PA laser is fixed at J = 1. (b) PI spectrum of molecules in the X$^{1}$$\Sigma$$^{+}$ (v'' = 0) state. In the following experiments, the frequency of PI laser is fixed at the transition of 2$^{1}$$\Pi$$_{1}$ (v' = 12) $\leftarrow$ X$^{1}$$\Sigma$$^{+}$ (v'' = 0). (c) Mass spectrum in two conditions: ``ODT off'' and ``ODT on''.}
  \label{f2}
\end{figure}

\subsection*{Loading procedure of optical trapped molecules}

The dynamical processes of $^{85}$Rb$^{133}$Cs molecules in the ODT will be analyzed in two different procedures: loading and holding. The loading procedure of molecules in the X$^{1}$$\Sigma$$^{+}$ (v'' = 0) state is observed [see Fig. 3(a)]. As we can see, the number of molecules in the X$^{1}$$\Sigma$$^{+}$ (v'' = 0) state loaded into the ODT increases rapidly until reaching a maximum value, which is followed by a number decrease due to a loss process.

A phenomenological rate equation is introduced to describe dynamical processes of molecules in the loading procedure~{\color{blue}\cite{KuppensCorwinMillerEtAl2000}}. The molecule number $N_{mol}$(t) in the loading procedure can be described as
\begin{equation}
\begin{split}
&\frac{dN_{mol}(t)}{dt}\\
&= R_{L}e^{-\gamma t}-\Gamma_{L}N_{mol}(t)-\beta_{L}\int_{V}n^{2}(r,t)dr^{3}. \label{eq:1} \\
\end{split}
\end{equation}

Here $R_{L}$ is the maximum loading rate when the molecules are loaded into ODT at the initial time, $\gamma$ represents the loss rate of molecules, $\Gamma _{L}$ denotes the single molecular collision with both background gas and co-trapped cold samples ($^{85}$Rb atoms, $^{133}$Cs atoms) in the dark-SPOTs, while $\beta_{L}$ is the loss rate of molecule-molecule cold collisions.~The subscript $\textit{L}$ is used to distinguish the loss rates during loading process from the holding process in the ODT.~Since the number of trapped molecules is much smaller (see texts below) than the trapped atoms, while their volume are nearly the same {\color{blue}\cite{DeiglmayrReppDulieuEtAl2011}}, the molecular density is much lower than the atomic density. Considering the universal inelastic rates for Rb-RbCs, Cs-RbCs, and RbCs-RbCs inelastic collisions are at the same order {\color{blue}\cite{DeiglmayrReppWesterEtAl2011}}, the $\beta_{L}$ is ignorable compared to  the Rb-RbCs, Cs-RbCs inelastic loss rates. Thus the Eq. (1) can be reduced as

\begin{equation}\label{2}
\frac{dN_{mol}(t)}{dt} = R_{L}e^{-\gamma t}-\Gamma_{L}N_{mol}(t)
\end{equation}

The analytical solution to this equation is

\begin{equation}\label{3}
\begin{split}
&N_{mol}(t)\\
&= \frac{e^{-\Gamma_{L}t}(\Gamma_{L}N_{0}-R_{L}+e^{(\Gamma_{L}-\gamma)t}R_{L}-N_{0}\gamma)}{\Gamma_{L}-\gamma}.\\
\end{split}
\end{equation}
Here, $N_{0}$ is the number of molecules loaded into ODT when the ODT beams are turned on.
Based on this solution, it is expected that the number of molecules increases linearly with the initial loading time, $N_{mol}(t)=R_{L}t$.  This is consistent with our observation. We choose the initial 15 ms to fit the maximum loading rate R$_{L}$ = 47 (2) s$^{-1}$ ions per pulse while the other parameters in the loading procedure are determined by fitting the measured ion signal to Eq. (3) (shown in the inset of Fig. 3(a)).

Based on Eq. (3), if $\gamma$ is zero, the number of molecules will gradually increase to a maximum value.~The $\Gamma_{L }$ and $\gamma$ are fitted to be 75~(4)~s$^{-1}$ and 1.4 (2) s$^{-1}$ ions per pulse with the experimental data.~The presence of $\gamma$ causes the decrease of the molecule number in loading procedure.~The measured number of $^{85}$Rb and $^{133}$Cs atoms in PA region, respectively, is not affected by the presence of ODT, which can exclude the possibility that the loss rate is induced by the decrease of atom number in the PA region. We then measure the number of trapped Rb and Cs atoms in ODT during the loading procedure as both atoms and molecules are loaded into ODT simultaneously. The results are shown in Fig. 3(b) and (c). The fitting curves are also based on Eq. (1), now modified for atom number $N_{atom}$. The number of atoms increases until reaching a maximum value. This is because the $\gamma_{atom}$, which represents the rate of the MOT loses atoms~{\color{blue}\cite{KuppensCorwinMillerEtAl2000}}, is zero because that there is no operation for dark-SPOT lasers in our experiment. We observe that the number of trapped molecules and atoms reach maximum values at nearly the same time, although their dynamics subsequent are quite different. Thus we attribute the loss rate $\gamma_{atom}$ to the loss rate of RbCs molecules in dark-SPOTs, as induced by the trapped atoms in ODT.

\begin{figure}[H]
 \setlength{\abovecaptionskip}{0cm}
 \centerline{\includegraphics[width=0.8\columnwidth]{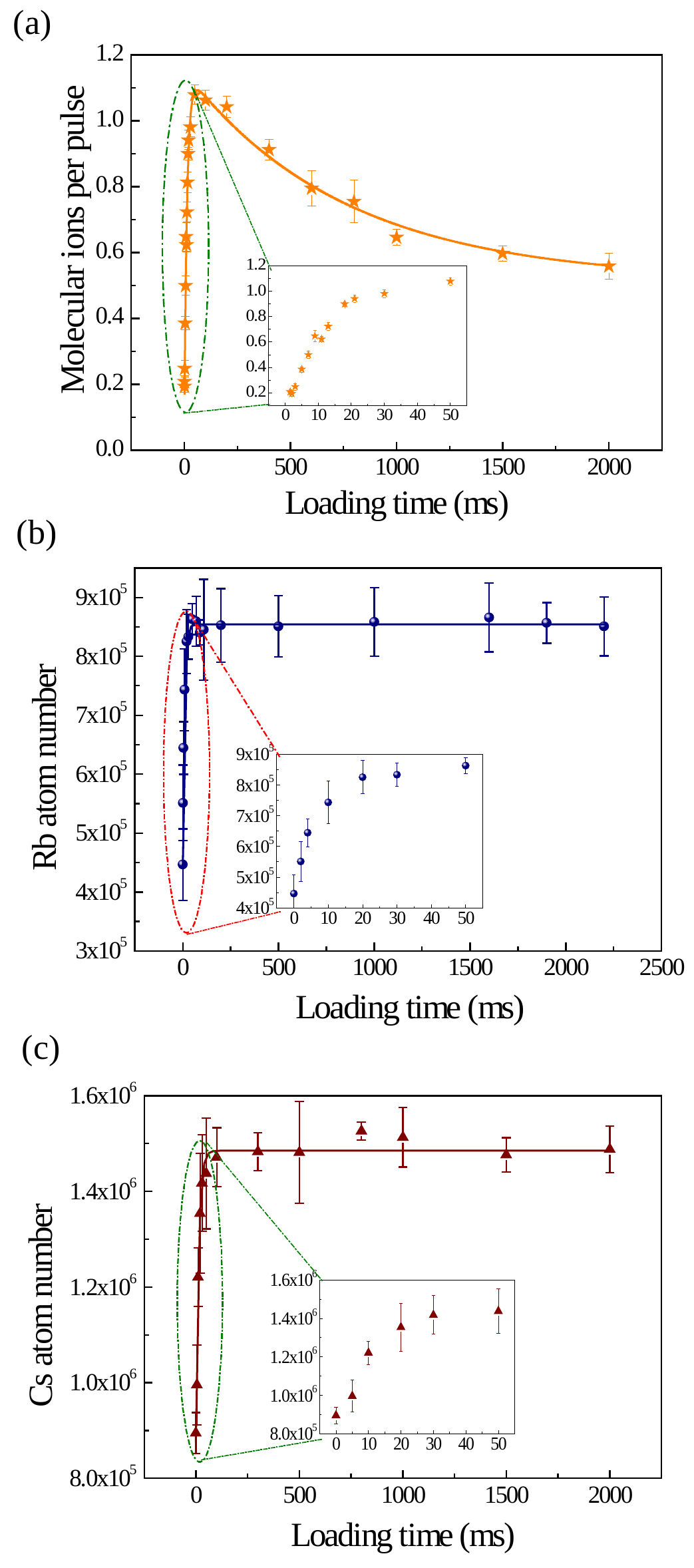}}
  \caption{ODT loading procedures of (a) $^{85}$Rb$^{133}$Cs molecules in the X$^{1}$$\Sigma$$^{+}$ (v'' = 0) state, (b) $^{85}$Rb atoms, and (c) $^{133}$Cs atoms. All the data are recorded at holding time T$_{H}$ = 1 ms. Insets: the molecule and atom number in initial time.}
  \label{f3}
\end{figure}

\subsection*{Holding procedure of optical trapped molecules}

First, we compare the time evolution of the molecule number in ``ODT off'' and  ``ODT on''. As shown in Fig.~4(a), the molecule number decreases quickly and has shorter lifetime in ``ODT off'' compared to the``ODT on'' case. The starting point in ``ODT off'' is the beginning of release procedure and the one in ``ODT on'' is the beginning of ODT holding procedure. Because the untrapped samples can escape from the region of detection and almost can not be observed after the release,~the trapped sample could only be observed during the holding procedure.~The trapped molecules loaded into ODT are the ones at the beginning of the release procedure and the number is fitted to be about 1.2~(1) ions per pulse (dashed line shown in Fig. 4(a)) according to the measurement of ``ODT on''. Considering the initial number in dark-SPOTs is about 6.0 (5) ions per pulse, the transfer efficiency of X$^{1}$$\Sigma$$^{+}$ (v'' = 0) state molecules from dark-SPOTs to ODT is estimated as 20 (2)\%.~The transit time in the RETPI beam is measured to be about 10 ms, the ionization efficiency and the detection efficiency of MCP are estimated to be around 5\%~{\color{blue}~\cite{C.Drag}} and 50\%~{\color{blue}\cite{Altaf2015}}, respectively. Thus the number of trapped molecules in X$^{1}$$\Sigma$$^{+}$ (v'' = 0) state is 48~(4), and the production rate of the X$^{1}$$\Sigma$$^{+}$ (v'' = 0) state molecules in ODT can be estimated as 4800~(400) /s.

The time evolution of molecule number in holding procedure can be described as~{\color{blue}\cite{NoViDis}}
\begin{equation}\label{4}
\begin{split}
&\frac{dN_{mol}(t)}{dt}\\
&=-\Gamma_{BG}N_{mol}(t)-\frac{1}{\sqrt{(1+q_{\alpha})^{3}}}\Gamma_{atom}N_{mol}(t)\\
&-\beta_{H}\int_{V}n^{2}(r,t)dr^{3}.\\
\end{split}
\end{equation}
Here, $\Gamma_{BG}$ is the loss rate of molecules due to its collision with the background gas, $\Gamma_{atom}$ is the loss rate of inelastic co-trapped atom-molecule collisions, $\beta_{H}$ is the loss rate of molecule-molecule cold collisions. The factor $\sqrt{(1+q_{\alpha})^{3}}$ takes into account of the different trap sizes for atoms and molecules, $q_{\alpha}$ is the ratio between the atomic and molecular polarizabilities~{\color{blue}\cite{DeiglmayrReppWesterEtAl2011}}. In the following analysis, the atomic polarizabilities are experimental values ($^{85}$Rb is 318.4 a.u.~{\color{blue}\cite{MolofSchwartzMillerEtAl1974}} and $^{133}$Cs  is 401.2 a.u.~{\color{blue}\cite{AminiGould2003}}). The molecular polarizability in X$^{1}$$\Sigma$$^{+}$ (v'' = 0) state is 597.6 a.u., which is supported by explicit \textit{ab initio} calculations~{\color{blue}\cite{DeiglmayrAymarWesterEtAl2008}}. As mentioned before, the $\beta_{H}$ can be ignored. So the Eq. (4) is simplified as
\begin{equation}\label{5}
\begin{split}
\frac{dN_{mol}(t)}{dt}&=-(\Gamma_{BG}+\frac{1}{\sqrt{(1+q_{\alpha})^{3}}}\Gamma_{atom})N_{mol}(t)\\
                      &=-\Gamma N_{mol}(t).\\
\end{split}
\end{equation}
This equation can be solved as
\begin{equation}\label{6}
 N_{mol}(t)=N_{0}e^{-t/\tau}
\end{equation}
Here, $\tau$ is the typical lifetime and $\tau^{-1}$ = $\Gamma$. The typical lifetime $\tau_{off}$  is 10 (5) ms and the molecules almost disappear after the release time, while the $\tau_{on}$ is 19 (4) ms after the release time. In ``ODT off'' condition, the molecules are not trapped and surrounded by atoms prepared in dark-SPOTs, so the $\Gamma_{off}$ comes from processes involving the molecule-background gas collisions, atom-molecule inelastic collisions and the dissipation process of thermal motion. Under ``ODT on'' condition, the molecules have been confined in the ODT, and have longer lifetime.~But,~the molecular lifetime and number are still limited by the collisions, including the atom-molecule inelastic collisions.

\begin{figure}[h]
  \centerline{ \includegraphics[width=0.8\columnwidth]{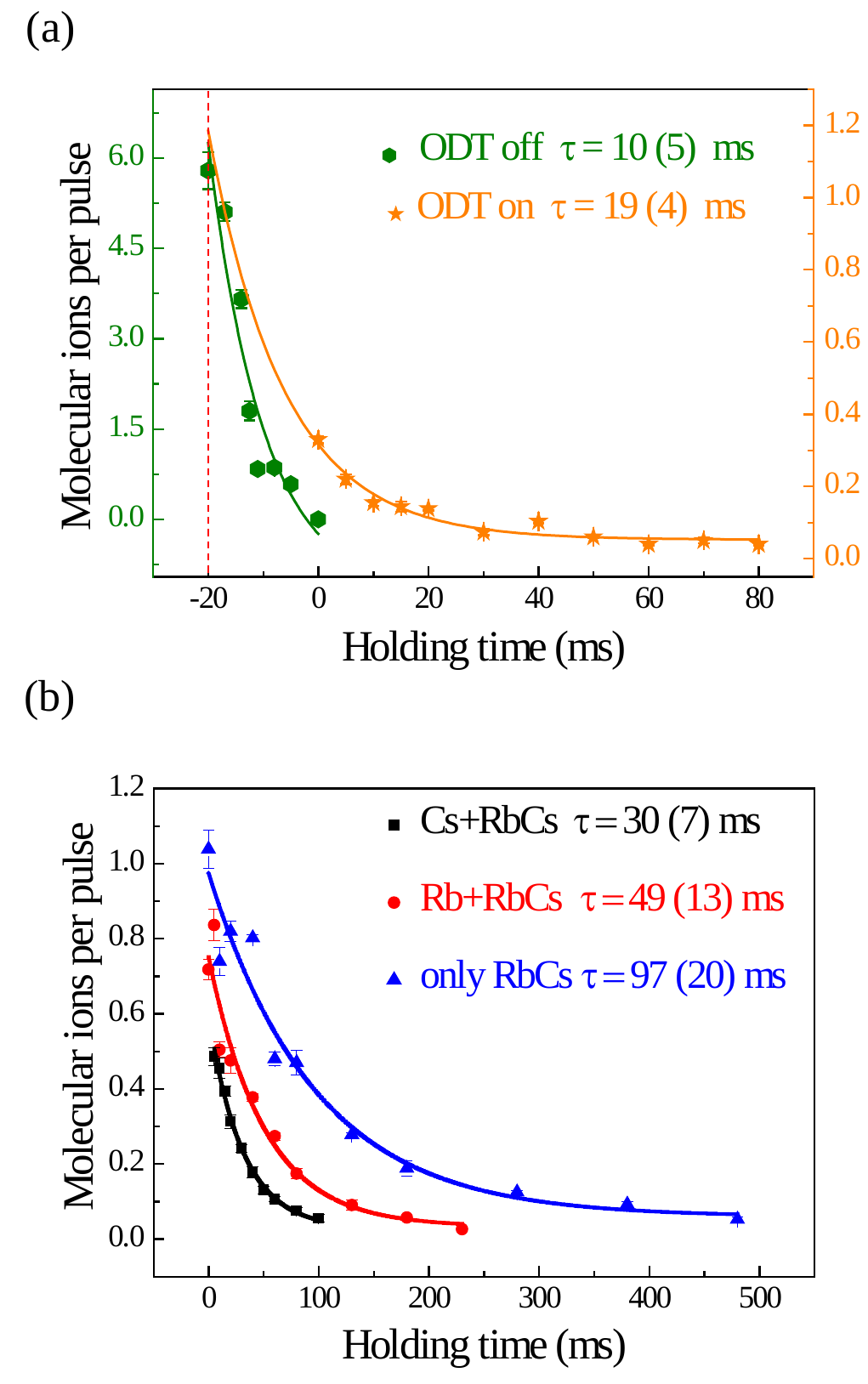}}
  \caption{Holding procedure of $^{85}$Rb$^{133}$Cs molecules in the X$^{1}$$\Sigma$$^{+}$ (v'' = 0) state. (a) the time evolution of  X$^{1}$$\Sigma$$^{+}$ (v'' = 0) state molecules in ``ODT off'' and ``ODT on''. (b) the time evolution of X$^{1}$$\Sigma$$^{+}$ (v'' = 0) state molecules in holding procedure in the absence of co-trapped atoms, and presence of Rb and Cs atoms, respectively.}
  \label{f4}
\end{figure}

To remove influences of atom-molecule inelastic collisions, the pushing lasers are adopted in our experiment. By selectively turning on the pushing laser, the  molecules or the molecules with chosen atoms could be prepared in ODT. Figure \ref{f4}(b) shows the time evolution of the $^{85}$Rb$^{133}$Cs molecules, the molecules with $^{85}$Rb atoms in F = 2 state and $^{133}$Cs atoms in F = 3 state in the holding procedure. As we can see, the absence of cold atoms in the ODT significantly increases the number and lifetime of the trapped molecules.

The introduction of pushing lasers allows us to observe the atom-molecule collisions. In molecular sample in X$^{1}$$\Sigma$$^{+}$ (v'' = 0) state in the trap, the typical lifetime is 97 (20) ms, which is consistent with the background gas-limited lifetime for isolated $^{133}$Cs atomic clouds in our measurement. These results also mean the molecular lifetime is limited by collisions with background gas. In this condition, the $\Gamma_{atom}$ = 0, $\tau^{-1}_{1} = \Gamma_{BG}$ =~10 (3) s$^{-1}$ by fitting with Eq. (6). So, the $\Gamma_{^{85}Rb}$ is deduced to be 19 (4) s$^{-1}$ from the data of molecules trapped with $^{85}$Rb atoms, and the $\Gamma_{^{133}Cs}$ is deduced to be~50 (12)~s$^{-1}$ from the data of molecules trapped with $^{133}$Cs atoms. The sum of $\Gamma_{RbCs}$ , $\Gamma_{^{85}Rb}$  and $\Gamma_{^{133}Cs}$  are 79 (8) s$^{-1}$,  agreeing with the value of  $\Gamma_{L}$. $\Gamma_{atom}$ is related to the inelastic collision rate K and $\Gamma_{atom}$ = Kn$_{atom }$. Assuming the atomic density is a constant, we extract the K$_{^{85}Rb}$ as 1.0 (2) $\times$10$^{-10}$ cm$^{3}$s$^{-1}$, K$_{^{133}Cs}$ is 1.2 (3)$ \times$ 10$^{-10}$ cm$^{3}$s$^{-1}$ from our data.~The released energy from the rovibrational quenching will remove the reactive atoms and molecules from the trap, thus the inelastic collisions of $^{85}$Rb$^{133}$Cs molecules in X$^{1}$$\Sigma$$^{+}$ (v'' = 0) state with $^{85}$Rb ($^{133}$Cs) atoms are not affected by the inelastic collisions of $^{85}$Rb$^{133}$Cs molecules in X$^{1}$$\Sigma$$^{+}$ (v'' > 0) states with $^{85}$Rb ($^{133}$Cs) atoms.~These values are on the same order as those reported in other experiments performed with trapped molecules in a$^{3}\Sigma^{+}$ states and  high lying level of X$^{1}$$\Sigma$$^{+}$ states, such as the $^{85}$Rb$^{133}$Cs molecules in a$^{3}\Sigma^{+}$ states, $^{7}$Li$^{133}$Cs molecules in a$^{3}\Sigma^{+}$ states and high lying level of X$^{1}$$\Sigma$$^{+}$ states~{\color{blue}\cite{HudsonGilfoyKotochigovaEtAl2008,DeiglmayrReppWesterEtAl2011}}. Besides, our inelastic rate is still higher than the computed one~{\color{blue}\cite{DeiglmayrReppWesterEtAl2011}}, which indicates that partial waves higher than the $\emph{s}$-wave have to be involved in the collision under our experimental conditions. We notice that although the produced molecules populate in several vibrational states, the measured results should be the same as the case for pure X$^{1}$$\Sigma$$^{+}$ (v'' = 0) state ones, due to ignorable vibrational redistribution and molecule-molecule cold collisions.

\section*{Conclusion}

We have shown that ultracold $^{85}$Rb$^{133}$Cs molecules in the lowest vibrational X$^{1}$$\Sigma$$^{+}$ (v'' = 0) ground state (to be precise, the J'' = 1 and J'' = 3 states of X$^{1}$$\Sigma$$^{+}$ (v'' = 0)) produced via short-range PA have been confined in a crossed ODT. The loading and holding procedures of X$^{1}$$\Sigma$$^{+}$ (v'' = 0) state molecules in ODT are analyzed based on a phenomenological rate equation. The inelastic collisions of $^{85}$Rb$^{133}$Cs molecules in the X$^{1}$$\Sigma$$^{+}$ (v'' = 0, J'' = 1 and J'' = 3) states with ultracold $^{85}$Rb (or $^{133}$Cs) atoms are measured. We note that ref.~{\color{blue}\cite{StwalleyBanerjeeBellosEtAl2010}} has suggested that the 2$^{3}$$\Pi$$_{0^{+}}$ state may have vibrational levels that permit efficient X$^{1}$$\Sigma$$^{+}$ (v'' = 0) state production in LiK, LiRb, LiCs, NaK, NaRb, NaCs and KCs systems.~So, our demonstration provides a simple and generic procedure for studying the dynamical process of trapped cold molecules in singlet ground states.

\section*{Acknowledgments}
We thank Y. Hu, Y. Yang and C. Li for the meaningful discussions. The work is supported by National Key Research and Development program (No.~2017YFA0304203), Natural Science Foundation of China (Nos.~61675120,~11434007 and~61378015), NSFC Project for Excellent Research Team (No.~61121064), Shanxi Scholarship Council of China, ``1331 KSC", and PCSIRT(No.~IRT~13076), Applied Basic Research Project of Shanxi Province (No.~201601D202008).


\scriptsize{
\bibliographystyle{rsc} } 

\end{document}